\documentclass[preprint]{article}
\usepackage{graphicx}% Include Figure files
\usepackage{dcolumn}% Align table columns on decimal point
\usepackage{bm}% bold math
\usepackage{amsmath}
\usepackage{amssymb}
\usepackage{multirow}
\usepackage{caption}
\usepackage{subcaption}
\usepackage{epstopdf}
\usepackage{hyperref}
\begin{document}
{\setlength{\oddsidemargin}{1.2in}
\setlength{\evensidemargin}{1.2in} } \baselineskip 0.55cm
\begin{center}
{\LARGE {\bf Barrow Holographic Dark Energy in $f(Q,L_{m})$ gravity: A dynamical system perspective}}
\end{center}
\date{\today}
\begin{center}
  Amit Samaddar, S. Surendra Singh \\
   Department of Mathematics, National Institute of Technology Manipur,\\ Imphal-795004,India\\
   Email:{ samaddaramit4@gmail.com, ssuren.mu@gmail.com }\\
 \end{center}
 
 \textbf{Abstract}: In this work, we investigate the cosmological implications of the Barrow Holographic Dark Energy (BADE) model within the framework of $f(Q,L_{m})$ gravity, specifically considering the model $f(Q,L_{m})=\alpha Q+\beta L_{m}$. Using a dynamical system approach for both non-interacting and interacting scenarios, we identify critical points corresponding to different phases of the Universe's evolution, including matter domination, radiation domination and dark energy-driven accelerated expansion. Our analysis reveals two stable critical points in the non-interacting case and three stable critical points in the interacting case, each indicating a transition to a stable phase dominated by BADE. The phase plots clearly demonstrate the evolution of the Universe's dynamics toward these stable points. At these stable points, the deceleration parameter is negative, consistent with accelerated expansion and the equation of state parameter suggests that BADE behaves as a dark energy component. These findings highlight the BADE model's strength as a viable explanation for the Universe's late-time acceleration inside $f(Q, L_{m})$ gravity, and they provide novel perspectives on the cosmic development of dark energy-matter interactions.
 
 
 \section{Introduction}\label{sec1}
 \hspace{0.5cm} The Universe, a vast and dynamic entity, has captivated human curiosity for centuries. The cosmos, which is made up of everything from the biggest galaxy clusters to the tiniest subatomic particles, is shaped by fundamental physical principles that dictate its structure and evolution. Modern cosmology, grounded in the framework of the Big Bang theory, posits that the Universe began as a hot, dense singularity approximately $13.8$ billion years ago. Since then, it has expanded, cooled and evolved into the complex cosmos, we observe today. In its present phase, the Universe is experiencing an accelerated expansion, a phenomenon first discovered through observations of distant Type Ia supernovae in the late $1990$s \cite{I99}. This discovery, which earned the $2011$ Nobel Prize in Physics, has led to the widespread acceptance of the $\Lambda$CDM (Lambda Cold Dark Matter) model. In this model, the Universe consists of approximately $68\%$ dark energy, $27\%$ dark matter and $5\%$ ordinary (baryonic) matter. The mysterious dark energy component is responsible for the current acceleration, while dark matter plays a crucial role in the formation of large-scale structures, such as galaxies and clusters. The rate of expansion of the universe, with recent measurements yielding values around $67.4$ km/s/Mpc from the Planck satellite's observations of the cosmic microwave background (CMB) \cite{Planck20} and approximately $73.0$ km/s/Mpc from the distance ladder approach using Cepheid variables and Type Ia supernovae \cite{Riess19,S99}. This discrepancy is known as the Hubble tension. A dimensionless measure of the Universe's acceleration, where negative values indicate acceleration. The current value is approximately $0.285\pm0.021$, reflecting the accelerated expansion. Despite the success of the $\Lambda$CDM model, certain puzzles remain unsolved, such as the nature of dark energy and the origin of the Hubble tension. These challenges have prompted the exploration of alternative cosmological models, including modifications to General Relativity (GR).
 
 Modified gravity theories (MGTs) have emerged as a crucial area of research in modern cosmology, offering potential explanations for the Universe's observed acceleration during both its early inflationary period and its current accelerated expansion. These theories go beyond Einstein's General Relativity (GR) by altering the fundamental geometric structure of spacetime or introducing new fields, thereby addressing cosmological phenomena that GR struggles to explain without invoking dark energy or dark matter. One of the earliest and most well-known modifications is $f(R)$ gravity, which extends the Einstein-Hilbert action by replacing the Ricci scalar $R$ with a general function $f(R)$ \cite{Nojiri11}. This theory has been extensively studied for its ability to drive cosmic inflation and late-time acceleration, with significant contributions by various researchers who have explored its cosmological implications \cite{P10,SM04}. Beyond $f(R)$ gravity, other extensions have been developed to include more complex couplings between matter and geometry. A prominent example is $f(R,L_{m})$ gravity introduced by \cite{Lobo10}, where the gravitational Lagrangian is a function of both the Ricci scalar $R$ and the matter-Lagrangian $L_{m}$. This approach has been explored for its potential to explain dark energy and the cosmic acceleration while also allowing for a more direct interaction between matter and the gravitational field. Studies have shown that this theory can lead to interesting astrophysical and cosmological consequences, such as modifications to the standard evolution of the universe and potential new insights into the nature of dark matter and dark energy \cite{S24, Y24}. Further extending the idea of modifying gravity, researchers have explored theories that alter the underlying geometric framework of GR itself. For instance, teleparallel gravity, known as $f(T)$ gravity, reformulates GR using torsion instead of curvature, providing an equivalent description of gravitational interactions, which is also called teleparallel equivalent of general relativity (TEGR) \cite{C61,C63}. Another significant advancement in this direction is $f(G)$ gravity, where $G$ is the Gauss-Bonnet invariant \cite{M11}. In addition, $f(R,T)$ gravity, where $T$ represents the trace of the energy-momentum tensor, allows for a coupling between matter and geometry that can account for deviations from GR at both cosmological and astrophysical scales \cite{T11}. More recently, the symmetric teleparallel equivalent of GR, known as $f(Q)$ gravity has been developed \cite{J20}. In this framework, the fundamental geometric variable is the non-metricity scalar $Q$, which geometrically defines the variation in the length of a vector during parallel transport. This theory has gained attention due to its novel approach to defining the gravitational interaction and has been applied to various cosmological models, including those addressing cosmic acceleration and dark energy \cite{W21,R19,F21}. Building upon $f(Q)$ gravity, researchers have introduced $f(Q,T)$ gravity, where the non-metricity scalar $Q$ is non-minimally coupled to the trace $T$ of the energy-momentum tensor. This theory further generalizes the interaction between matter and geometry, providing new avenues for understanding the Universe's accelerated expansion \cite{Harko20}.
 
 Building upon the $f(Q)$ framework and the concept of non-minimal coupling, \cite{Hazarika24} introduced $f(Q,L_{m})$ gravity, which incorporates the matter Lagrangian into the Lagrangian density of $f(Q)$ gravity, leading to the formulation of $f(Q,L_{m})$ gravity. This theory allows for both minimal and non-minimal couplings between geometry and matter, providing a flexible framework for exploring the interplay between matter fields and the underlying geometry of spacetime. The authors derived the general field equations and explored the conservation laws within this theory, finding that the matter energy-momentum tensor is generally not conserved, indicating potential new physics beyond standard GR. They also investigated the cosmological evolution under the flat Friedmann-Lema$\hat{i}$tre-Robertson-Walker (FLRW) metric, deriving the generalized Friedmann equations and exploring specific models that demonstrate the rich phenomenology of $f(Q,L_{m})$ gravity. In $f(Q,L_{m})$ gravity, the analytical solutions and observational analysis are discussed in \cite{Y24}. \cite{M24} analyzed the behavior of bulk viscosity in $f(Q,L_{m})$ gravity.

 Barrow Holographic Dark Energy (BADE) represents a novel approach to addressing the dark energy problem, building upon the foundational principles of holography and fractal geometry. Traditional holographic dark energy (HDE) models are grounded in the holographic principle, which suggests that the energy content of the Universe can be linked to its horizon entropy \cite{S17}. This entropy follows the classical Bekenstein-Hawking entropy formulation, which is typically related to black holes and is proportional to the area of the horizon \cite{Barrow20}. Inspired by John Barrow's work, BADE distinguishes itself by adding a fractal variation to this entropy-area connection. In this extended model, the horizon entropy is no longer merely proportional to the horizon area but is instead modified by a fractal parameter $\Delta$, which quantifies the deviation from the standard entropy formula. This modification introduces a new layer of complexity and flexibility into the holographic dark energy framework, allowing it to better accommodate various cosmological observations \cite{E20}. Moreover, the BADE model has been investigated in several cosmological situations, including those associated with early universe inflation, the evolution of the equation of state (EoS) parameter and the model's compatibility with other theoretical frameworks like cyclic cosmology and braneworld scenarios \cite{Ghosh24,A24}. These studies highlight the versatility of BADE in addressing the dark energy problem and suggest that it can be a powerful tool in exploring the Universe's late-time dynamics. Some authors have recently studied the Barrow holographic dark energy models in various gravity theories \cite{A21,Dixit21,A22,H22}.

 The analysis of the intricate evolution of the cosmos can be facilitated by the use of dynamical systems, which are essential to cosmology. The dynamical system approach allows cosmologists to study the behavior of cosmological models by converting the equations of motion, often derived from Einstein's field equations or modified gravity theories, into a set of first-order differential equations. These equations describe the evolution of key cosmological parameters, such as the Hubble parameter, energy densities and scale factor, in terms of dimensionless variables \cite{F12}. One of the primary benefits of the dynamical system approach is its ability to uncover the long-term behavior of cosmological models without requiring explicit solutions to the differential equations. Rather than this, the method concentrates on finding critical points, which are associated with steady-state solutions where the system's parameters either stay constant or change predictably. These critical points represent different cosmic stages, like matter-dominated, radiation-dominated, or dark energy-dominated periods. The stability analysis of these critical points is particularly valuable, as it provides insight into the nature of the Universe's evolution. Stable critical points indicate attractor solutions, towards which the universe naturally evolves, while unstable points suggest scenarios where small perturbations can lead to drastically different outcomes \cite{W18}. This stability study provides insights into the ultimate destiny of the Universe and aids in comprehending its dynamics in the past, present and future. In recent years, a wide range of modified gravity theories have been applied using the dynamical system technique, including $f(R)$, $f(Q)$, $f(Q,T)$, $f(T)$, $f(T,B)$ gravity theories and others, as well as to dark energy models like quintessence, phantom energy and holographic dark energy \cite{ Mishra22, C23, Mishra23, LD24, Franco20, Mirza16}. For models like BADE in the context of $f(Q,L_{m})$ gravity, the dynamical system approach allows for a detailed examination of the interplay between dark energy and the underlying gravitational theory. It contributes to our understanding of how the expansion of the Universe is affected by the fractal modification of entropy in BADE and whether a more realistic account of cosmic acceleration can be obtained using the modified gravity framework.

 The layout of this paper is given as follows: In Section \ref{sec2}, we derive the field equations for $f(Q,L_{m})$ gravity, providing the foundational equations that govern the dynamics within this theoretical framework. Section \ref{sec3} delves into the dynamical system approach for BADE in $f(Q,L_{m})$ gravity. We perform a detailed stability analysis of the critical points and present the corresponding phase portraits. Finally, in Section \ref{sec4}, we summarize our findings and discuss the broader implications of our results.
 
 \section{Field equations and theoretical framework of $f(Q,L_{m})$ gravity}\label{sec2}
 \hspace{0.5cm} In this section, we discuss the theoretical aspects of $f(Q,L_{m})$ gravity and derive its field equations. When a metric is established, the Riemann tensor provides a geometric interpretation of gravity as well as its contractions which is
\begin{equation}\label{1}
R^{a}_{b\mu\nu}=\partial_{\mu}Y^{a}_{\nu b}-\partial_{\nu}Y^{a}_{\mu b}+Y^{a}_{\mu\psi} Y^{\psi}_{\nu b}-Y^{a}_{\nu\psi} Y^{\psi}_{\mu b},
\end{equation}
Through the use of an affine connection, the Riemann tensor is built. Torsion and non-metricity are two features of Weyl-Cartan geometry, which is a development of Riemannian geometry. Under this structure, the affine connection $Y^{a}_{\mu\nu}$ may be broken down into three independent parts: the disformation tensor $L^{a}_{\mu\nu}$, the symmetric Levi-Civita connection $\Gamma^{a}_{\mu\nu}$ and contortion tensor $K^{a}_{\mu\nu}$. Therefore, we can formulate it as follows \cite{Y19}:
\begin{equation}\label{2}
Y^{a}_{\mu\nu}=\Gamma^{a}_{\mu\nu}+L^{a}_{\mu\nu}+K^{a}_{\mu\nu}.
\end{equation}
The Levi-Civita connection $\Gamma^{a}_{\mu\nu}$ is a fundamental concept in differential geometry, representing a torsion-free connection that preserves the metric $g_{\mu\nu}$. It is fully described by the metric and its first derivatives, governing the curvature and parallel transport within the context of general relativity, thus capturing the essence of gravitational interactions.
\begin{equation}\label{3}
\Gamma^{a}_{\mu\nu}=\frac{1}{2}g^{a\psi}(\partial_{\mu}g_{\psi\nu}+\partial_{\nu}g_{\psi\mu}-\partial_{\psi}g_{\mu\nu}),
\end{equation}
The contortion tensor $K^{a}_{\mu\nu}$ is a geometric object that quantifies the deviation from a torsion-free connection. It captures the effects of torsion in spacetime and is defined in terms of the difference between the affine connection $Y^{a}_{\mu\nu}$ and the Levi-Civita connection $\Gamma^{a}_{\mu\nu}$. The expression for the contortion tensor is given by:
\begin{equation}\label{4}
K^{a}_{\mu\nu}=\frac{1}{2}(T^{a}_{\mu\nu}+T_\mu{}^{a}{}_{\nu}+T_\nu{}^{a}{}_{\mu}),
\end{equation}
where $T^{a}_{\mu\nu}$ is the torsion tensor. The torsion tensor, which captures spacetime's intrinsic twisting, yields the torsion scalar $T$. It gives a measurement of the departure from a torsion-free geometry and is defined as the trace of the contraction of the torsion tensor with itself. In torsion-based theories of gravity, the torsion scalar is an important factor that affects the equations controlling the spacetime manifold's dynamics.

The deformation tensor $L^{a}_{\mu\nu}$ captures the effects of non-metricity in a connection, reflecting how vector lengths may vary during parallel transport. It measures the deviation from metric compatibility, indicating that the metric tensor is not preserved along parallel transport paths. This tensor is crucial in theories where the geometric structure extends beyond Riemannian frameworks, encompassing more general spacetime geometries. It is defined as:
\begin{equation}\label{5}
L^{a}_{\mu\nu}=\frac{1}{2}(Q^{a}_{\mu\nu}-Q_\mu{}^{a}{}_{\nu}-Q_\nu{}^{a}{}_{\mu}),
\end{equation}
where $Q^{a}_{\mu\nu}$ represents the non-metricity tensor, capturing the deviation of the metric from being covariantly constant. The definition of the non-metricity tensor is given by
\begin{equation}\label{6}
Q_{a\mu\nu}=\nabla_{a}g_{\mu\nu}=\partial_{a}g_{\mu\nu}-Y^{b}_{a\mu}g_{b\nu}-Y^{b}_{a\nu}g_{b\mu},
\end{equation}

An introduction of the superpotential tensor $P^{a}_{\mu\nu}$, a conjugate of the non-metricity, is required to establish a boundary component in the action of metric-affine gravitational theories. This superpotential is defined as a tensor that relates to the non-metricity tensor $Q^{a}_{\mu\nu}$ and is crucial for constructing boundary terms that ensure the action's consistency and the preservation of variational principles. The superpotential helps in capturing the contributions from the non-metricity in a manner that aligns with the geometric and physical properties of the theory. The superpotential tensor $P^{a}_{\mu\nu}$ is defined as,
\begin{equation}\label{7}
P^{a}_{\mu\nu}=-\frac{1}{2}L^{a}_{\mu\nu}+\frac{1}{4}(Q^{a}-\widetilde{Q}^{a})g_{\mu\nu}-\frac{1}{4}\delta^{a}{}_{(\mu}Q_{\nu)},
\end{equation}
where $Q^{a}$=$Q^{a}{}_{\mu}{}^{\mu}$ and $\widetilde{Q}^{a}$=$Q_{\mu}{}^{a\mu}$ depict the vectors of non-metricity. One can derive the non-metricity scalar by contracting the non-metricity tensor with the superpotential tensor as
\begin{equation}\label{8}
Q=-Q_{\psi\mu\nu}P^{\psi\mu\nu}.
\end{equation}
The non-metricity scalar $Q$ quantifies the extent to which the geometry of a manifold deviates from being purely Riemannian. It serves as an indicator of how the shape or orientation of an object changes during parallel transport, independent of torsion. In particular, $Q$ measures the failure of the metric to remain constant when an object is moved through spacetime, reflecting the influence of non-metricity on the overall structure of the manifold.

In the $f(Q,L_{m})$ theory, the gravitational action is expressed as follows:
\begin{equation}\label{9}
S=\int f(Q,L_{m})\sqrt{-g}d^{4}x,
\end{equation}
In this expression: $(i)$ $S$ represents the total action of the gravitational system, $(ii)$ $f(Q,L_{m})$ is a general function of the non-metricity scalar $Q$ and the matter Lagrangian $L_{m}$. This function encapsulates the dynamics of both the gravitational field and the matter content of the Universe, $(iii)$ $\sqrt{-g}$ is the square root of the negative determinant of the metric tensor $g_{\mu\nu}$, which ensures that the action is invariant under coordinate transformations and properly accounts for the volume element in curved spacetime and $(iv)$ $d^{4}x$ denotes the four-dimensional volume element, integrating over the entire spacetime manifold.

By changing the action (\ref{9}) with respect to the metric tensor $g_{µv}$, the field equation are derived as follows:
\begin{equation}\label{10}
\frac{2}{\sqrt{-g}}\nabla_{a}(f_{Q}\sqrt{-g}P^{a}_{\mu\nu})+f_{Q}(P_{\mu a\beta}Q_{\nu}{}^{a\beta}-2Q^{a\beta}{}_{\mu}P_{a\beta\nu})+\frac{1}{2}fg_{\mu\nu}=\frac{1}{2}f_{L_{m}}(g_{\mu\nu}L_{m}-T_{\mu\nu}),
\end{equation}
where $f_{Q}=\frac{\partial f}{\partial Q}$ and $f_{L_{m}}=\frac{\partial f}{\partial L_{m}}$. The operator $\nabla_{a}$ is the covariant derivative. $T_{\mu\nu}$ is the stress-energy tensor, which describes the distribution of matter and energy in spacetime. The stress-energy-momentum tensor is typically defined as:
\begin{equation}\label{11}
T_{\mu\nu}=-\frac{2}{\sqrt{-g}}\frac{\delta(\sqrt{-g}L_{m})}{\delta g^{\mu\nu}}=g_{\mu\nu}L_{m}-2\frac{\partial L_{m}}{\partial g^{\mu\nu}},
\end{equation}
The field equations can be obtained by altering the gravitational action concerning the connection as follows:
\begin{equation}\label{12}
\nabla_{\mu}\nabla_{\nu}\bigg[4\sqrt{-g}f_{Q}P^{\mu\nu}_{a}+H^{\mu\nu}_{a}\bigg]=0,
\end{equation}
Here, $H^{\mu\nu}$ represents the density of hypermomentum, which includes spin, dilation and shear contributions from the matter fields. It generalizes the notion of stress-energy to account for more complex matter properties. The expression of $H^{\mu\nu}$ is as follows:
\begin{equation}\label{13}
H^{\mu\nu}=\sqrt{-g}f_{L_{m}}\frac{\delta L_{m}}{\delta Y^{a}_{\mu\nu}}.
\end{equation}
Through the implementation of the covariant derivative to the field equation (\ref{10}), we can find,
\begin{equation}\label{14}
D_{\mu}T^{\mu}_{\nu}=\frac{1}{f_{L_{m}}}\bigg(\frac{2}{\sqrt{-g}}\nabla_{a}\nabla_{\mu}H^{a\mu}_{\nu}+\nabla_{\mu}A^{\mu}_{\nu}-\nabla_{\mu}\bigg[\frac{1}{\sqrt{-g}}\nabla_{a}H^{a\mu}_{\nu}\bigg]\bigg    )=B_{\nu}\neq 0.
\end{equation}
Thus, from equation (\ref{14}), in $f(Q,L_{m})$ gravity, the matter energy-momentum tensor $T_{\mu\nu}$ is generally not conserved. This non-conservation is characterized by a tensor $B_{\nu}$ which is contingent upon the dynamic variables $Q$ and $L_{m}$ and thermodynamic parameters. The presence of $B_{\nu}$ indicates that the interaction between geometry and matter leads to a deviation from the standard conservation laws seen in general relativity. The equation $D_{\mu}T^{\mu}_{\nu}$=$B_{\nu}\neq 0$ reflects this non-conservation, highlighting the modified dynamics in this theory.

To explore the cosmological implications of $f(Q,L_{m})$ gravity, we consider the Friedmann-Lema$\hat{i}$tre-Robertson-Walker (FLRW) metric in Cartesian coordinates, which describes a homogeneous and isotropic universe. The FLRW metric in Cartesian coordinates is given by:
\begin{equation}\label{15}
ds^{2}=-dt^{2}+a^{2}(t)(dx^{2}+dy^{2}+dz^{2}),
\end{equation}
Here, the scale factor $a(t)$ is a function of cosmic time and represents the relative expansion (or contraction) of the universe. It scales the spatial coordinates and determines how distances between objects change over time. In the framework of the FLRW metric, the non-metricity scalar $Q$ is expressed as $Q=6H^{2}$, where $H=\frac{\dot{a}}{a}$ is the Hubble parameter. This scalar quantifies the deviation of the manifold's geometry from metric preservation and reflects how the rate of expansion of the universe affects the non-metricity. The Hubble parameter $H$ measures the rate at which the scale factor $a(t)$ changes over time and thus $Q$ directly links the geometric properties of the Universe to its dynamic expansion behavior.

For the FLRW metric, which represents a universe filled with perfect fluid matter, the energy-momentum tensor $T_{\mu\nu}$  is defined to encapsulate the properties of this matter content. It is expressed as:
\begin{equation}\label{16}
T_{\mu\nu}=(p+\rho)u_{\mu}u^{\nu}+pg_{\mu\nu},
\end{equation}
where $\rho$  is the energy density, $p$ represents the pressure of the fluid and $u_{\mu}$ denotes the four-velocity of the fluid.

In the context of $f(Q,L_{m})$ gravity, the evolution of an FLRW universe with matter represented as a perfect fluid is governed by modified Friedmann equations. The modified Friedmann equations are given by:
 \begin{equation}\label{17}
3H^{2}=\frac{1}{4f_{Q}}\bigg[f-f_{L_{m}}(\rho+L_{m})\bigg],
 \end{equation}
 \begin{equation}\label{18}
\dot{H}+3H^{2}+\frac{\dot{f_{Q}}}{f_{Q}}H=\frac{1}{4f_{Q}}\bigg[f+f_{L_{m}}(p-L_{m})\bigg].
 \end{equation}
 where the dot notation represents the time derivative of a quantity. The goal of developing these equations is to provide an accurate mathematical explanation of the interplay between matter and geometry in this system. These equations serve as the basis for investigating a wide range of cosmological and astrophysical scenarios, including the evolution of large-scale structures, the behaviour of cosmic acceleration and the interplay between dark matter and dark energy. The field equations also enable us to compare the $f(Q,L_{m})$ model's viability to empirical evidence, which offers crucial insights into prospective changes to our understanding of gravity.
 \section{Dynamical system analysis in BADE for $f(Q,L_{m})$ gravity}\label{sec3}
 \subsection{Dynamical system}\label{sec3.1}
\hspace{0.5cm} Dynamical systems theory offers a profound framework for understanding the evolution of various physical systems, including cosmological models. At its core, a dynamical system is a set of functions that describe how the state of a system evolves over time according to specific rules, typically represented by differential equations. The roots of dynamical systems analysis can be traced back to the early work of Sir Isaac Newton, who formulated the laws of motion that govern the gravitational interactions in systems like the sun and planets. While Newton's work provided exact solutions for the two-body problem, it was Henri Poincar$\acute{e}$ in the late $19$th century who revolutionized the field by introducing qualitative methods to analyze more complex systems, such as the three-body problem, where no general analytical solution exists.

In cosmology, dynamical systems analysis has become an indispensable tool for exploring the behavior of the universe at different epochs. The cosmological evolution can be described by a set of ordinary differential equations (ODEs) derived from the Einstein field equations, where the variables represent key physical quantities such as the scale factor, Hubble parameter and energy densities of different components (radiation, matter, dark energy, etc.) \cite{Sonia22}. Specifically, an ODE takes the form $\dot{u}=h(u)$, where $\dot{u}=\frac{du}{dt}$, $u=(u_{1},u_{2},u_{3},......,u_{m})$ is a vector of variables in $\mathbb{R}^{m}$, $u:\mathbb{R}^{m}\rightarrow \mathbb{R}^{m}$ and $h(u)$ represents the governing equations of the system \cite{Amit23}. If these equations do not explicitly depend on time, the system is said to be autonomous, which simplifies the analysis and interpretation. The phase space of a dynamical system is the multi-dimensional space where each point represents a possible state of the system, defined by the values of the variables $u_{1},u_{2},u_{3},......,u_{m}$ \cite{Singh19}. The trajectories in phase space, known as phase paths, illustrate how the system evolves over time from one state to another. A key concept in dynamical systems analysis is the identification of equilibrium points, where the system's state remains unchanged over time, i.e., $h(u)=0$ \cite{Singh23}. These points are critical for understanding the long-term behavior of the system and can be classified based on their stability properties. The stability of an equilibrium point is determined by examining the system's response to small perturbations. If, after a perturbation, the system returns to the equilibrium point, the point is said to be stable or an attractor. Conversely, if the system moves away from the equilibrium point, it is unstable or repulsive. An equilibrium point with a mix of stable and unstable directions is known as a saddle point \cite{Amit2}. Furthermore, equilibrium points can be categorized as hyperbolic or non-hyperbolic depending on the eigenvalues of the Jacobian matrix associated with the system. A hyperbolic equilibrium point has no zero eigenvalues, while a non-hyperbolic one has at least one zero eigenvalue, indicating more complex behavior \cite{Amit24,Rathore23}.

In the context of cosmology, understanding the stability of equilibrium points is crucial for determining the behavior of the Universe at different stages. For example, during the inflationary epoch, the universe undergoes rapid expansion driven by a repulsive force, which should correspond to an unstable equilibrium point in a cosmological model. As the Universe transitions through radiation and matter-dominated eras, these phases should be represented by stable equilibrium points. Finally, the current accelerated expansion, driven by dark energy, should act as a stable attractor in the phase space, ensuring that the Universe will continue to expand indefinitely \cite{Boehmer85}.

\subsection{Barrow holographic dark energy (BADE)}\label{sec3.2}
\hspace{0.5cm} Barrow introduced an innovative concept that extends the classical view of black hole entropy by incorporating quantum-gravitational effects. In classical thermodynamics, the entropy $S'$ of a black hole is directly proportional to the area $A$ of its event horizon, a relationship known as the Bekenstein-Hawking entropy. However, Barrow proposed a generalized form of entropy, taking into account possible quantum corrections, which can be expressed as:
\begin{equation}\label{19}
S_{B}=\bigg(\frac{A}{A_{0}}\bigg)^{1+\frac{\Delta}{2}},
\end{equation}
Here, $A$ is the area of the black hole horizon, $A_{0}$ is the Planck area and $\Delta$ is a parameter that quantifies the deviation from the standard entropy due to quantum-gravitational deformations. When $\Delta=0$, the usual Bekenstein-Hawking entropy is recovered, indicating no quantum corrections. For $\Delta>0$, the entropy increases, reflecting the complexity introduced by quantum effects.

Building on this generalized entropy, Barrow proposed a modification to the conventional holographic dark energy model, resulting in what is known as Barrow Holographic Dark Energy (BADE). In the standard holographic dark energy model, the energy density $\rho$ is linked to the inverse square of a cosmological length scale, typically the Hubble horizon or future event horizon. Barrow's modification introduces a dependence on the parameter $\Delta$, leading to a new expression for the energy density \cite{EN20}:
\begin{equation}\label{20}
\rho_{BD}=CL^{\Delta-2},
\end{equation}
where $C$ is a constant with dimensions dependent on $\Delta$ and $L$ represents a characteristic length scale, such as the Hubble scale or the future event horizon. The parameter $\Delta$ governs the deviation from the standard holographic energy density, with $\Delta=0$ yielding the traditional holographic dark energy model. The dimension of $C$ is $[L]^{-\Delta-2}$. A commonly chosen IR cutoff in cosmology is the conformal time of the Universe. When the conformal time $\eta$ is used as the IR cutoff, the energy density for BADE is expressed as \cite{U21}:
\begin{equation}\label{21}
\rho_{BD}=C\eta^{\Delta-2},
\end{equation}
Here, $\eta$ is the conformal time, which is related to the scale factor $a$ and the Hubble parameter $H$ through the integral:
\begin{equation}\label{22}
\eta=\int_{0}^{a}\frac{da}{Ha^{2}},
\end{equation}
The conformal time $\eta$ serves as a measure of the ``age" of the universe when scaled by the expansion factor, providing a natural choice for the IR cutoff. The relation $dt=ad\eta$ connects the proper time $t$ and the conformal time $\eta$, where $dt$ is the differential proper time. This approach introduces a dependency of the dark energy density on the conformal time, leading to a more generalized model that incorporates quantum-gravitational effects through the Barrow parameter $\Delta$. This flexibility allows the model to better fit observational data and provides a more comprehensive description of the Universe's late-time acceleration.

\subsection{Interpretation of the dynamical system for BADE with linear $f(Q,L_{m})$ model} \label{sec3.3}
\hspace{0.5cm} In this section, we explore a linear model within the framework of $f(Q,L_{m})$ gravity, where the function $f(Q,L_{m})$ is taken as a linear combination of the non-metricity scalar $Q$ and the matter lagrangian $L_{m}$. The linear model is expressed as:
\begin{equation}\label{23}
f(Q,L_{m})=\alpha Q+\beta L_{m},
\end{equation}
Here, $\alpha$ and $\beta$ are constants that represent the coupling strengths between the non-metricity scalar $Q$ and the matter lagrangian $L_{m}$. Comparing to non-linear models, this choice simplifies the gravitational action and results in simple field equations that are easier to analyze. The model allows for the direct interaction between the geometric part, represented by $Q$, and the matter content of the universe through $L_{m}$. For specific choices of $\alpha$ and $\beta$, the model can reduce to well-known cases. For example, setting $\alpha=1$ and $\beta=0$ recovers the teleparallel equivalent of general relativity, whereas $\alpha=0$ and $\beta=1$ yield a model purely dependent on the matter Lagrangian.

 In the context of our linear model, we have chosen to identify the matter Lagrangian as $L_{m}=\rho$ \cite{Harko15}. This assumption simplifies the model by reducing the complexity associated with the matter Lagrangian. It allows us to focus on the energy density, a well-understood quantity in cosmology, making the equations more tractable and easier to interpret. With $L_{m}=\rho$, the energy density plays a dual role: it is both the source of gravitational fields and a driver of the Universe's expansion. This dual role is critical in understanding how the Universe transitions between different phases, such as from matter domination to dark energy domination.

 To analyze the cosmological dynamics, we introduce the following dimensionless variables from equation (\ref{17}) as:
 \begin{equation}\label{24}
 x=\frac{\rho_{r}}{3H^{2}}, \hspace{0.5cm} y=\frac{\rho_{BD}}{3H^{2}}, \hspace{0.5cm} z=\frac{f}{3H^{2}},
 \end{equation}
 where the total energy density of the Universe is given by the sum of the contributions from matter, radiation and BADE:
 \begin{equation}\label{25}
\rho=\rho_{m}+\rho_{r}+\rho_{BD},
 \end{equation}
Here, $\rho_{m}$ is the energy density of matter and $\rho_{r}$ represents the energy density of radiation. The corresponding pressures are defined as follows: $(i)$ For matter (non-relativistic), the pressure is $p_{m}=0$, $(ii)$ For radiation, the pressure is $p_{r}=\frac{\rho_{r}}{3}$, consistent with the equation of state for radiation and $(iii)$ For BADE, the pressure is $p_{BD}=\omega_{BD}\rho_{BD}$. The energy conservation equations are:
\begin{equation}\label{26}
\dot{\rho_{m}}+3H\rho_{m}=0, \hspace{0.5cm} \dot{\rho_{r}}+4H\rho_{r}=0, \hspace{0.5cm} \dot{\rho}_{BD}+3H\rho_{BD}(1+\omega_{BD})=0.
\end{equation}
The equation of state (EoS) parameter $\omega_{BD}$ for BADE is crucial in characterizing the nature of this dark energy component. It is defined as:
\begin{equation}\label{27}
\omega_{BD}=\frac{\rho_{BD}}{p_{BD}},
\end{equation}
From equations (\ref{21}) and (\ref{26}), we derive the expression of $\omega_{BD}$ as follows:
\begin{equation}\label{28}
\omega_{BD}=-1-\frac{\Delta-2}{3},
\end{equation}
With the help of the dimensionless variables (\ref{24}), the density parameters are derived as follows from the equation (\ref{17}):
\begin{equation}\label{29}
\Omega_{r}=x, \hspace{0.5cm} \Omega_{BD}=y, \hspace{0.5cm} \Omega_{m}=-\frac{2\alpha}{\beta}+\frac{z}{2\beta}-x-y.
\end{equation}
The expression of $\frac{\dot{H}}{H^{2}}$ are obtained from the equations (\ref{17}) and (\ref{18}) as:
\begin{equation}\label{30}
\frac{\dot{H}}{H^{2}}=-\frac{3}{2}-\frac{3z}{4\alpha}+\frac{\beta x}{4\alpha}-\frac{\beta y}{4\alpha}(\Delta+1),
\end{equation}
The expression (\ref{30}) indicates that the evolution of the Hubble parameter is influenced by a balance between different cosmic components matter, radiation and BADE , each contributing differently to the overall dynamics. The interaction of these components through the parameters $x$, $y$ and $z$ along with the constants $\alpha$ and $\beta$, determines whether the Universe experiences deceleration or acceleration at different stages of its evolution.

The deceleration parameter helps to understand the phase of the Universe's expansion, indicating how different cosmic components influence whether the Universe is accelerating or decelerating at any given time. Using the expression of $\frac{\dot{H}}{H^{2}}$, we obtain the deceleration parameter $(q)$ as:
\begin{equation}\label{31}
q=-1-\frac{\dot{H}}{H^{2}}=\frac{1}{2}+\frac{3z}{4\alpha}-\frac{\beta x}{4\alpha}+\frac{\beta y}{4\alpha}(\Delta+1).
\end{equation}
The overall behavior of $q$ depends on the terms involving $x$, $y$, $z$ and the parameters $\alpha$, $\beta$ and $\Delta$.

Now, we present the autonomous dynamical system equations derived from the dimensionless variables defined earlier. These equations encapsulate the evolution of the Universe under the influence of various cosmological components, including matter, radiation and BADE with in the context of $f(Q,L_{m})$ gravity. By substituting the dimensionless variables into the field equations and using the assumptions specific to our model, we obtain the following autonomous system of first-order differential equations:
\begin{equation}\label{32}
\frac{dx}{dN}=-2x\bigg[\frac{1}{2}-\frac{3z}{4\alpha}+\frac{\beta x}{4\alpha}-\frac{\beta y}{4\alpha}(\Delta+1)\bigg],
\end{equation}
\begin{equation}\label{33}
\frac{dy}{dN}=(\Delta-2)y-2y\bigg[-\frac{3}{2}-\frac{3z}{4\alpha}+\frac{\beta x}{4\alpha}-\frac{\beta y}{4\alpha}(\Delta+1)\bigg],
\end{equation}
\begin{equation}\label{34}
\frac{dz}{dN}=\frac{3z}{2\alpha}(z-\alpha)-\frac{\beta xz}{2\alpha}+\frac{\beta yz}{2\alpha}(\Delta+1).
\end{equation}
where $N=loga$ is the e-folding number. Having derived the autonomous dynamical system equations for our model, we now proceed to analyze the critical points and their stability. To find the critical points of the system of equations (\ref{32})-(\ref{34}), we solve the equations $\frac{dx}{dN}$=$\frac{dy}{dN}$=$\frac{dz}{dN}$=$0$. We calculate the eigenvalues from the Jacobian matrix at each critical point to analyze the stability criteria of our model. The table exhibits the critical points along with their corresponding cosmological parameters as outlined in Table \ref{Tab:T1}. The stability conditions and characteristic values are depicted in Table \ref{Tab:T2}. We obtain five critical points for the above system of equations.\\
\begin{table}[h!]
\centering
\caption{Critical points $\&$ the physical parameters for the system of equations (\ref{32})-(\ref{34})}
\begin{tabular}{||p{0.9cm}|p{1.9cm}|p{1.0cm}|p{2.4cm}|p{0.7cm}|p{1.0cm}|p{2.8cm}||}
\hline\hline
 Points &\hspace{0.7cm} $x$ &\hspace{0.4cm} $y$ & \hspace{0.9cm} $z$ & $\hspace{0.3cm}q$ & $\hspace{0.2cm}\omega_{BD}$& Exists for\\
\hline\hline
$\hspace{0.3cm}A$ & \hspace{0.8cm}$0$ & \hspace{0.5cm}$0$ & $\hspace{1.0cm}0$ & \hspace{0.3cm}$\frac{1}{2}$ & \hspace{0.2cm}$\approx0$ & $\Delta=0.1<1$\\[1pt]
\hline
$\hspace{0.3cm}B$ & $y(\Delta+1)-\frac{2\alpha}{\beta}$ & \hspace{0.5cm}$y$ & $\hspace{1.0cm}0$ & \hspace{0.3cm}$1$ & \hspace{0.1cm}$-0.33$ & $y\neq 0$\\[1pt]
\hline
$\hspace{0.3cm}C$ & \hspace{0.8cm}$0$ & $\frac{-6\alpha}{\beta(\Delta+1)}$ & $\hspace{1.0cm}0$ & $-1$ & $\frac{-\Delta-1}{3}$ & $\beta\neq 0$\\[1pt]
\hline
$\hspace{0.3cm}D$ & \hspace{0.2cm}$-\frac{2\alpha}{\beta}+\frac{3z}{\beta}$ & \hspace{0.5cm}$0$ & \hspace{1.0cm}$z$ & \hspace{0.3cm}$1$ & \hspace{0.1cm}$-0.33$ & $z\neq 0$, $\beta\neq 0$, $0<\Delta<1$\\[1pt]
\hline
$\hspace{0.3cm}E$ & \hspace{0.8cm}$0$ & \hspace{0.5cm}$y$ & $-2\alpha-\frac{\beta y}{3}(\Delta+1)$ & $-1$ & $\frac{-\Delta-1}{3}$ & $y\neq 0$, $0<\Delta<1$ \\[1pt]
\hline\hline
\end{tabular}
\label{Tab:T1}
\end{table}
\begin{table}[h!]
\centering
\caption{Eigenvalues and their stability criteria}
\begin{tabular}{||p{0.9cm}|p{1.8cm}|p{2.5cm}|p{2.2cm}|p{4.2cm}||}
\hline\hline
 Points &\hspace{0.5cm} $\gamma_{1}$ &\hspace{0.8cm} $\gamma_{2}$ & \hspace{0.6cm} $\gamma_{3}$ & Criteria \\
\hline\hline
$\hspace{0.3cm}A$ & \hspace{0.4cm}$-1$ & \hspace{0.5cm}$\Delta+1$ & $\hspace{0.5cm}-\frac{3}{2}$ & saddle for $\Delta=0.1$\\
\hline
$\hspace{0.3cm}B$ & $1-\frac{\beta y(\Delta+1)}{2\alpha}$ & \hspace{0.4cm}$\Delta-\frac{\beta y(\Delta+1)}{2\alpha}$ & $\hspace{0.5cm}-\frac{1}{2}$ & saddle for $y<\frac{2\alpha}{\beta(\Delta+1)}$\\[1pt]
\hline
$\hspace{0.3cm}C$ & \hspace{0.4cm}$-4$ & \hspace{0.5cm}$\Delta-5$ & $\hspace{0.5cm}-\frac{9}{2}$ & stable for $0<\Delta<1$\\[1pt]
\hline
$\hspace{0.3cm}D$ & \hspace{0.2cm}$1-\frac{3z}{2\alpha}$ & \hspace{0.9cm}$\Delta$ & $-\frac{1}{2}+\frac{3z}{2\alpha}$ & unstable for $\frac{\alpha}{3}<z<\frac{2\alpha}{3}$ $\&$ $\Delta>0$\\[1pt]
\hline
$\hspace{0.3cm}E$ & \hspace{0.4cm}$-\frac{9}{2}$ & $\Delta-2+\frac{\beta y}{2\alpha}(\Delta+1)$ & $-7-\frac{\beta y}{\alpha}(\Delta+1)$ & stable for $y<\frac{2\alpha}{\beta}\frac{2-\Delta}{(\Delta+1)}$ $\&$ $0<\Delta<1$\\[1pt]
\hline\hline
\end{tabular}
\label{Tab:T2}
\end{table}
\begin{figure}[hbt!]
    \centering
    \begin{subfigure}[b]{0.3\textwidth}
        \includegraphics[scale=0.4]{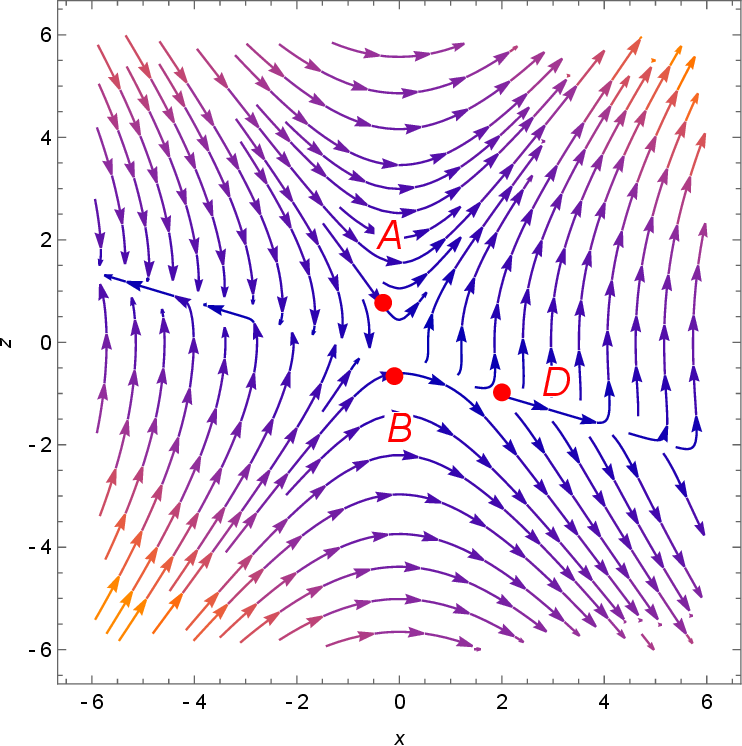}
        \caption{}
        \label{fig:subfig1}
    \end{subfigure}
    \hfill
    \begin{subfigure}[b]{0.3\textwidth}
        \includegraphics[scale=0.4]{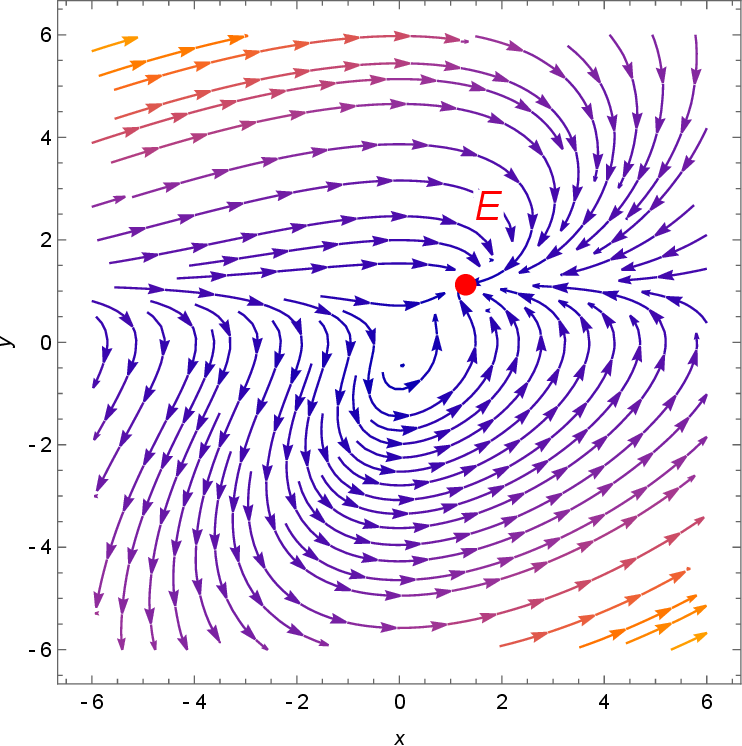}
        \caption{}
        \label{fig:subfig2}
    \end{subfigure}
    \hfill
    \begin{subfigure}[b]{0.3\textwidth}
        \includegraphics[scale=0.4]{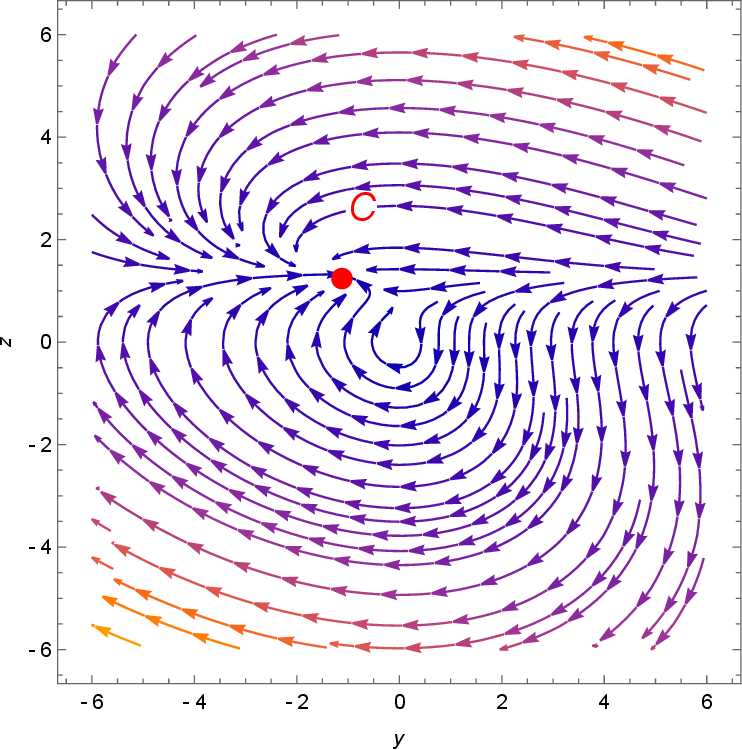}
        \caption{}
        \label{fig:subfig3}
    \end{subfigure}

    \caption{Phase space diagram for the equations (\ref{32})-(\ref{34}) with $\alpha=-1$, $\beta=2$ and $\Delta=0.1$.}
    \label{fig:f1}
\end{figure}
\begin{table}[h!]
\centering
\caption{Physical behavior of the scale factor at each equilibrium points}
\begin{tabular}{||p{0.9cm}|p{3.5cm}|p{3.0cm}|p{3.3cm}||}
\hline\hline
 Points & Accelerating equation & \hspace{0.4cm}Scale factor & Phase \\
\hline\hline
$\hspace{0.3cm}A$ & \hspace{0.6cm}$\dot{H}=-\frac{3}{2}H^{2}$ & $a(t)=a_{0}(\frac{3t}{2}+b_{1})^{\frac{2}{3}}$ & matter dominated\\
\hline
$\hspace{0.3cm}B$ & \hspace{0.6cm}$\dot{H}=-2H^{2}$ & $a(t)=a_{0}(2t+b_{2})^{\frac{1}{2}}$ & radiation dominated\\
\hline
$\hspace{0.3cm}C$ & \hspace{0.8cm}$\dot{H}=0$ & \hspace{0.5cm}$a(t)=a_{0}e^{b_{3}t}$ & de-Sitter\\
\hline
$\hspace{0.3cm}D$ & \hspace{0.6cm}$\dot{H}=-2H^{2}$ & $a(t)=a_{0}(2t+b_{2})^{\frac{1}{2}}$ & radiation dominated\\
\hline
$\hspace{0.3cm}E$ & \hspace{0.8cm}$\dot{H}=0$ & \hspace{0.5cm}$a(t)=a_{0}e^{b_{3}t}$ & de-Sitter\\
\hline\hline
\end{tabular}
\label{Tab:T3}
\end{table}\\

$\star$ \textbf{Critical Point $A$:} At the critical point $A$, $(x,y,z)=(0,0,0)$. The eigenvalues of the Jacobian matrix are $-1, \Delta+1, -\frac{3}{2}$. With the assumption $\Delta=0.1$, the eigenvalues become $-1, 1.1, -\frac{3}{2}$. The eigenvalues $-1, -\frac{3}{2}$ are negative, indicating stability in the corresponding directions. However, the eigenvalue $\Delta+1=1.1$ is positive, which indicates instability in that direction. This mix of stable and unstable directions implies that the critical point $A$ is a saddle point. In cosmological terms, this means that while the system might approach this state along certain trajectories, it will diverge away from it along others, showing that the Universe will not remain in this state permanently.

The deceleration parameter at this critical point is $q=\frac{1}{2}$, which corresponds to a Universe that is decelerating. This value of $q$ typically describes a matter-dominated phase where gravitational attraction slows down the expansion of the Universe. The equation of state parameter for BADE at this point is $\omega_{BD}=-0.3\approx 0$. This suggests that at this critical point, BADE behaves similarly to pressureless matter (dust), which does not contribute significantly to the acceleration or deceleration of the Universe. The density parameters at the critical point are $\Omega_{r}=\Omega_{BD}=0$ and $\Omega_{m}=1$. These values indicate that at this state, the Universe is entirely dominated by matter. The scale factor for point $A$ is expressed as $a(t)=a_{0}(\frac{3t}{2}+b_{1})^{\frac{2}{3}}$,$a_{0}$ represents scale factor at time $t=0$ which signifies the Universe's matter-dominated stage.

The plot \ref{fig:subfig1} of the dynamical system around this critical point would show trajectories approaching the point along the stable directions (corresponding to the negative eigenvalues) and diverging away from it along the unstable direction (corresponding to the positive eigenvalue). This visualization reinforces the interpretation of the Universe passing through but not remaining in the matter-dominated phase associated with this critical point.

$\star$ \textbf{Critical Point $B$:} At critical point $B$, the critical points are given by the coordinates $\bigg (y(\Delta+1)-\frac{2\alpha}{\beta},y,0\bigg)$, which exists for non-zero values of $y$. The eigenvalues associated with this point are $\big(1-\frac{\beta y(\Delta+1)}{2\alpha}, \Delta-\frac{\beta y(\Delta+1)}{2\alpha}, -\frac{1}{2}\big)$. The eigenvalues reveal important stability characteristics: The first eigenvalue $1-\frac{\beta y(\Delta+1)}{2\alpha}$  is positive if $y<\frac{2\alpha}{\beta(\Delta+1)}$, indicating instability in one direction. The second eigenvalue $\Delta-\frac{\beta y(\Delta+1)}{2\alpha}$ could be either positive or negative depending on the value of $y$, but for $y<\frac{2\alpha}{\beta(\Delta+1)}$, it is typically negative, implying stability in that direction. The third eigenvalue $-\frac{1}{2}$ is negative, confirming stability in this direction. Since one eigenvalue is positive and the others are negative, this critical point is characterized as a saddle point. This suggests that the Universe can approach this state under certain conditions but will eventually deviate from it.

The deceleration parameter at this point is $q=1$, which corresponds to a Universe in a state of deceleration. This is indicative of a phase where the Universe's expansion is being significantly slowed down, possibly due to the dominance of a particular component such as radiation. The equation of state parameter for BADE for the point $B$ is $\omega_{BD}=-0.33$. This value indicates that BADE behaves like a dark energy component with negative pressure, although it is less negative than a cosmological constant (where $\omega=-1$). This suggests that BADE contributes to the decelerating expansion, but not as strongly as a pure cosmological constant would. The density parameters are $\Omega_{m}=\Omega_{BD}=0$ and $\Omega_{r}=1$. These values indicate that at this critical point, the Universe is entirely dominated by radiation. The scale factor for this point is given by $a(t)=a_{0}(2t+b_{2})^{\frac{1}{2}}$, which describes a Universe where the scale factor evolves as the square root of time. This time-dependence is consistent with a radiation-dominated universe, where the scale factor typically grows as $a(t)\propto t^{\frac{1}{2}}$.

A plot around this critical point \ref{fig:subfig1} would show trajectories approaching it along the stable directions and diverging along the unstable direction. This behaviour is consistent with the hypothesis that the Universe may stabilize in a radiation-dominated phase for a brief period of time before changing into a different phase.

$\star$ \textbf{Critical Point $C$:} At critical point $C$, the coordinates of the critical point are $\bigg(0, \frac{-6\alpha}{\beta(\Delta+1)}, 0\bigg)$, with $\beta$ being non-zero. The eigenvalues associated with this critical point are $\big(-4, \Delta-5, -\frac{9}{2}\big)$. All eigenvalues are negative for $0<\Delta<1$, which indicates that this point is stable within this range of $\Delta$. The negative eigenvalues imply that any small perturbation around this critical point will decay, and the system will return to this state, making it an attractor in the phase space.

The deceleration parameter $q=-1$ corresponds to an accelerating Universe. This value of $q$ is characteristic of a de Sitter phase, where the expansion of the Universe is driven by a cosmological constant or a dark energy component that causes exponential growth in the scale factor. The equation of state parameter for BADE for the point $C$ is $\omega_{BD}=\frac{-\Delta-1}{3}$. This expression indicates that the BADE component behaves like a form of dark energy with a negative EoS parameter. For $\Delta$ within the range $0<\Delta<1$, the EoS parameter is less than $-\frac{1}{3}$, which is sufficient to drive cosmic acceleration, consistent with the observed behavior of dark energy. At this point $C$, the density parameters are $\Omega_{m}=\Omega_{r}=0$ and $\Omega_{BD}=1$. These values indicate that the Universe is entirely dominated by the BADE component. The absence of matter $(\Omega_{m}=0)$ and radiation $(\Omega_{r}=0)$ suggests that this point corresponds to a late-time phase of the Universe where dark energy has completely taken over, leading to an accelerated expansion.

The scale factor at the point $C$ is given by $a(t)=a_{0}e^{b_{3}t}$, which describes an exponentially expanding Universe. This behavior is characteristic of a de Sitter Universe, where the expansion rate is constant and the Universe undergoes a phase of eternal inflation driven by dark energy.

The stability of this critical point can be visualized in a phase space plot \ref{fig:subfig3} where trajectories converge towards this point, indicating its nature as a stable attractor. This reflects the idea that, regardless of initial conditions, the Universe is likely to evolve towards this accelerating phase dominated by dark energy.

$\star$ \textbf{Critical Point $D$:} At critical point $D$, $\big(-\frac{2\alpha}{\beta}+\frac{3z}{\beta},0,z\big)$, where $z\neq0$ and $\beta\neq0$. The eigenvalues associated with this critical point are $\big(1-\frac{3z}{2\alpha}, \Delta, -\frac{1}{2}+\frac{3z}{2\alpha}\big)$. The stability of this point is determined by the value of $z$ relative to $\alpha$ and $\Delta$. For $\frac{\alpha}{3}<z<\frac{2\alpha}{3}$ $\&$ $\Delta>0$, the point is unstable. Specifically, the eigenvalues indicate that perturbations along certain directions will grow, leading the system away from this critical point.

The deceleration parameter $q=1$ corresponds to a decelerating universe, typically associated with a radiation-dominated phase. This suggests that this critical point may represent an early stage in the Universe's evolution when radiation was the dominant component. The equation of state parameter for BADE for the point $D$ is $\omega_{BD}=-0.33$. Although this value indicates a dark energy-like behavior, the dominance of radiation at this point suggests that the BADE component is not significant enough to influence the overall dynamics at this stage. The density parameters are $\Omega_{m}=\Omega_{BD}=0$ and $\Omega_{r}=1$. This configuration implies that the Universe is entirely dominated by radiation. The scale factor at this critical point is given by $a(t)=a_{0}(2t+b_{2})^{\frac{1}{2}}$, which describes a Universe expanding with time $t$ in a manner characteristic of a radiation-dominated era. A signature of such a phase, where radiation is slowing down the expansion of the Universe, is the $t^{\frac{1}{2}}$ dependency.

The unstable behavior of this critical point can be visualized by plotting trajectories in the phase space \ref{fig:subfig1} that diverge from this point. The unstable nature of this point, particularly for $\frac{\alpha}{3}<z<\frac{2\alpha}{3}$ $\&$ $\Delta>0$, indicates that this phase is transient. The instability of this point $D$ suggests that the Universe will eventually move away from this radiation-dominated state as it evolves. As the radiation density decreases over time due to the expansion of the universe, the system will transition to other critical points corresponding to matter or dark energy domination.

$\star$ \textbf{Critical Point $E$:} The coordinates for the critical points $E$ are $\big(0,y,-2\alpha-\frac{\beta y}{3}(\Delta+1)\big)$. The eigenvalues relevant to this point $E$ are: $\big(-\frac{9}{2},\Delta-2+\frac{\beta y}{2\alpha}(\Delta+1),-7-\frac{\beta y}{\alpha}(\Delta+1)\big)$. The stability of this point hinges on the value of $y$ relative to $\alpha$, $\beta$ and $\Delta$. Specifically, for $y<\frac{2\alpha}{\beta}\frac{2-\Delta}{(\Delta+1)}$ and $0<\Delta<1$, the eigenvalues are all negative, indicating that this critical point is stable. This indicates that the Universe will attract towards this state when it is close to this point, acting as an attractor in the phase space.

 The deceleration parameter $q=-1$ signifies a Universe undergoing accelerated expansion. This value is typical of a phase dominated by dark energy, where the expansion of the Universe accelerates due to the negative pressure exerted by the dark energy component. The equation of state parameter for the point $E$ is $\omega_{BD}=\frac{-\Delta-1}{3}$ indicates that the BADE component behaves similarly to a cosmological constant or a quintessence-like field, depending on the value of $\Delta$. For $\Delta$ close to $1$, $\omega$ approaches the value corresponding to a cosmological constant $(\omega=-1)$ suggesting that this phase could represent a late-time dark energy-dominated era. The density parameters are $\Omega_{m}=\Omega_{r}=0$ and $\Omega_{BD}=1$. This implies that at this critical point, the Universe is completely dominated by the Barrow holographic dark energy, with no significant contribution from matter or radiation. Such a scenario aligns with the expectation for the Universe's future, where dark energy could become the sole component driving cosmic evolution. The scale factor $a(t)=a_{0}e^{b_{3}t}$ represents an exponential expansion. This exponential growth is consistent with a de Sitter-like phase, where the Universe's expansion accelerates without bound, leading to an ever-increasing scale factor.

 The parameter $\Delta$ plays a crucial role in determining the exact nature of this phase. For values of $\Delta$ close to $1$, the behavior closely mimics that of a cosmological constant, leading to a smooth, stable expansion. However, for smaller values of $\Delta$, the EoS parameter deviates from $-1$, potentially introducing variations in the expansion rate and the Universe's future dynamics.

  The stability plot \ref{fig:subfig2} for this point would show trajectories converging towards this critical point, reflecting its attractor nature. The plot would demonstrate that, regardless of initial conditions, the Universe is likely to settle into this accelerated expansion phase, highlighting the robustness of this cosmological scenario.
  \subsection{Dynamical system of BADE for interacting model in $f(Q,L_{m})$ gravity}\label{sec3.4}
  \hspace{0.5cm} In cosmological models, interactions between dark energy and other components, such as matter or radiation, are often considered to address issues like the cosmic coincidence problem. The energy exchange between various components can impact the evolution of the cosmos by adding an interaction term, $Q$, into the energy conservation equations. This interaction changesÂ the Universe's dynamics, which may have consequences for the late-time acceleration or even change the rate at which structures emerge. In our model, we consider an interaction term $Q=3H\gamma\rho_{BD}$ \cite{J13}, where $\gamma$ is a coupling constant that quantifies the strength of the interaction between BADE and other components. This interaction term suggests that the exchange of energy is proportional to the energy density of BADE and the expansion rate of the Universe. The sign of the interaction term plays a crucial role in determining the direction of energy transfer between the BADE and other components, such as matter. When $Q>0$, energy is transferred from the BADE component to the matter component. This scenario leads to an increase in the energy density of matter $(\rho_{m})$ and a corresponding decrease in the energy density of BADE $(\rho_{BD})$. When $Q<0$, energy is transferred from the matter component to the BADE component. In this case, the energy density of matter decreases faster than it would in the absence of interaction, while the energy density of BADE increases or depletes more slowly. A slower transition to a Universe dominated by dark energy may also be predicted by the model if $Q>0$, which could be consistent with some empirical evidence that points to a gradual shift. Conversely, if $Q<0$, the model might describe a more rapidly accelerating Universe, which could be consistent with observations indicating an increasing rate of cosmic expansion. The energy conservation equations for the interacting components in our model are modified by the interaction term $Q$. The following equations describe how the energy densities of matter, radiation, and dark energy evolve over time due to the interaction as,
  \begin{equation}\label{35}
 \dot{\rho_{m}}+3H\rho_{m}=Q, \hspace{0.5cm} \dot{\rho_{r}}+4H\rho_{r}=0, \hspace{0.5cm} \dot{\rho}_{BD}+3H(1+\omega_{BD})\rho_{BD}=-Q.
  \end{equation}
  We make the following dimensionless variable assumptions to execute the dynamical system for the interaction model:
  \begin{equation}\label{36}
  x=\frac{\rho_{r}}{3H^{2}}, \hspace{0.5cm} y=\frac{\rho_{BD}}{3H^{2}}, \hspace{0.5cm} z=\frac{f}{3H^{2}}, \hspace{0.5cm} r=\frac{\rho_{m}}{3H^{2}},
  \end{equation}
  From equations (\ref{21}) and (\ref{35}), we derive the EoS parameter's expression for interacting model as follows:
  \begin{equation}\label{37}
  \omega_{BD}=\frac{-\Delta-1}{3}+\gamma,
  \end{equation}
  Employing the equations (\ref{17}) and (\ref{18}), the expression of $\frac{\dot{H}}{H^{2}}$ in terms of the above variables (\ref{36}) is determined as follows:
  \begin{equation}\label{38}
  \frac{\dot{H}}{H^{2}}=\frac{3z}{4\alpha}-\frac{\beta x}{2\alpha}-\frac{\beta y}{4\alpha}(\Delta+4)+\frac{3\beta\gamma y}{4\alpha}-\frac{3\beta r}{4\alpha}-3,
  \end{equation}
 Substituting the expression of $\frac{\dot{H}}{H^{2}}$ from equation (\ref{38}) into the formula of the deceleration parameter, then the deceleration parameter $q$ becomes,
 \begin{equation}\label{39}
 q=2-\frac{3z}{4\alpha}+\frac{\beta x}{2\alpha}+\frac{\beta y}{4\alpha}(\Delta+4)-\frac{3\beta\gamma y}{4\alpha}+\frac{3\beta r}{4\alpha}.
 \end{equation}
 This expression encapsulates the effects of the various components, including radiation $(x)$, Barrow dark energy $(y)$, matter $(r)$ and the interaction term $Q$ on the cosmic deceleration. The resulting value of $q$ will determine whether the Universe is undergoing acceleration $(q<0)$ or deceleration $(q>0)$ in this interacting scenario.

 For the interacting case, the inclusion of an interaction term between BADE and matter introduces additional complexity into the previously derived autonomous dynamical system. To account for the interaction, we introduce a new dimensionless variable $r=\frac{\rho_{m}}{3H^{2}}$, representing the matter density. Thus, the system of first-order differential equations are given by:
 \begin{equation}\label{40}
 \frac{dx}{dN}=-4x-2x\bigg[\frac{3z}{4\alpha}-\frac{\beta x}{2\alpha}-\frac{\beta y}{4\alpha}(\Delta+4)+\frac{3\beta\gamma y}{4\alpha}-\frac{3\beta r}{4\alpha}-3\bigg],
 \end{equation}
 \begin{equation}\label{41}
 \frac{dy}{dN}=-6\gamma y-(2-\Delta)y-2y\bigg[\frac{3z}{4\alpha}-\frac{\beta x}{2\alpha}-\frac{\beta y}{4\alpha}(\Delta+4)+\frac{3\beta\gamma y}{4\alpha}-\frac{3\beta r}{4\alpha}-3\bigg],
 \end{equation}
 \begin{equation}\label{42}
 \frac{dz}{dN}=9z-6\beta x-6\beta y-6\beta r-12\alpha-\frac{3z^{2}}{2\alpha}+\frac{\beta xz}{\alpha}+\frac{\beta yz}{2\alpha}(\Delta+4)-\frac{3\beta\gamma yz}{2\alpha}+\frac{3\beta rz}{2\alpha},
 \end{equation}
 \begin{equation}\label{43}
 \frac{dr}{dN}=3\gamma y-3r-2r\bigg[\frac{3z}{4\alpha}-\frac{\beta x}{2\alpha}-\frac{\beta y}{4\alpha}(\Delta+4)+\frac{3\beta\gamma y}{4\alpha}-\frac{3\beta r}{4\alpha}-3\bigg].
 \end{equation}
 Here, $N=loga$ is the e-folding number, which tracks the expansion of the Universe. To find the critical points, we solve the system of equations $\frac{dx}{dN}=\frac{dy}{dN}=\frac{dz}{dN}=\frac{dr}{dN}=0$. The stability of these critical points is determined by analyzing the eigenvalues of the Jacobian matrix, which is derived from the system of equations (\ref{40})â€“(\ref{43}). The sign and magnitude of the eigenvalues indicate whether a particular critical point is stable (attractor), unstable (repeller), or a saddle point. The interaction parameter $\gamma$ plays a crucial role in modifying the stability conditions compared to the non-interacting case. The table of critical points, shown in Table \ref{Tab:T4}, lists the corresponding cosmological parameters, such as the deceleration parameter $q$ and the equation of state parameter $\omega_{BD}$. The stability analysis, summarized in Table \ref{Tab:T5}, reveals how the interaction influences the nature of these critical points.
 \begin{table}[h!]
\centering
\caption{Critical points $\&$ the physical parameters for the system of equations (\ref{40})â€“(\ref{43})}
\begin{tabular}{||p{0.9cm}|p{1.9cm}|p{3.5cm}|p{0.8cm}|p{1.0cm}|p{0.7cm}|p{1.0cm}|p{2.8cm}||}
\hline\hline
 Points &\hspace{0.7cm} $x$ &\hspace{1.5cm} $y$ & \hspace{0.2cm} $z$ & \hspace{0.4cm} $r$ & $\hspace{0.3cm}q$ & $\hspace{0.2cm}\omega_{BD}$& Exists for\\
\hline\hline
$\hspace{0.3cm}A_{1}$ & \hspace{0.8cm}$0$ & \hspace{1.6cm}$0$ & $\hspace{0.3cm}0$ & \hspace{0.2cm}$-\frac{2\alpha}{\beta}$ & \hspace{0.3cm}$\frac{1}{2}$ & \hspace{0.2cm}$\approx0$ & $\beta\neq0$\\[1pt]
\hline
$\hspace{0.3cm}B_{1}$ & \hspace{0.4cm}$-\frac{2\alpha}{\beta}$ & \hspace{1.6cm}$0$ & $\hspace{0.3cm}0$ & \hspace{0.5cm}$0$ & \hspace{0.3cm}$1$ & $-0.33$ & $\beta\neq0$\\[1pt]
\hline
$\hspace{0.3cm}C_{1}$ & \hspace{0.8cm}$0$ & \hspace{0.8cm}$\frac{-12\alpha}{\beta(\Delta+4-3\gamma)}$ & $\hspace{0.3cm}0$ & \hspace{0.5cm}$0$ & $-1$ & $\frac{-\Delta-1}{3}$ & $\beta\neq 0$, $0<\Delta<1$\\[1pt]
\hline
$\hspace{0.3cm}D_{1}$ & \hspace{0.1cm}$-\frac{2\alpha}{\beta}+\frac{3z}{2\beta}$ & \hspace{1.6cm}$0$ & \hspace{0.3cm}$z$ & \hspace{0.5cm}$0$ & \hspace{0.3cm}$1$ & $-0.33$ & $z\neq 0$, $\beta\neq 0$\\[1pt]
\hline
$\hspace{0.3cm}E_{1}$ & \hspace{0.8cm}$0$ & $\frac{1}{\Delta+4-3\gamma}\bigg[-\frac{12\alpha}{\beta}-3r\bigg]$ & $\hspace{0.3cm}0$ & \hspace{0.5cm}$r$ & $-1$ & $\frac{-\Delta-1}{3}$ & $r\neq 0$, $\beta\neq 0$, $0<\Delta<1$ \\[1pt]
\hline
$\hspace{0.3cm}F_{1}$ & \hspace{0.8cm}$0$ & $\frac{1}{\Delta+4-3\gamma}\bigg[-\frac{12\alpha}{\beta}+\frac{3z}{\beta}\bigg]$ & $\hspace{0.3cm}z$ & \hspace{0.5cm}$0$ & $-1$ & $\frac{-\Delta-1}{3}$ & $z\neq 0$, $\beta\neq 0$, $0<\Delta<1$ \\[1pt]
\hline
$\hspace{0.3cm}G_{1}$ & \hspace{0.1cm}$-\frac{3\alpha}{\beta}-\frac{3r}{2}$ & \hspace{1.6cm}$0$ & \hspace{0.3cm}$0$ & \hspace{0.5cm}$r$ & \hspace{0.3cm}$\frac{1}{2}$ & \hspace{0.2cm}$\approx0$ & $r\neq 0$, $\beta\neq 0$\\[1pt]
\hline\hline
\end{tabular}
\label{Tab:T4}
\end{table}
\begin{table}[h!]
\centering
\caption{Characteristic values and their stability criteria}
\begin{tabular}{||p{0.9cm}|p{1.3cm}|p{2.6cm}|p{1cm}|p{1.5cm}|p{3.9cm}||}
\hline\hline
 Points &\hspace{0.3cm} $\gamma_{1}$ &\hspace{0.8cm} $\gamma_{2}$ & \hspace{0.3cm} $\gamma_{3}$ & \hspace{0.4cm} $\gamma_{4}$ & Criteria \\
\hline\hline
$\hspace{0.3cm}A_{1}$ & \hspace{0.3cm}$-1$ & $-6\gamma+\Delta+7$ & $\hspace{0.5cm}6$ & $\hspace{0.4cm}-3$ & saddle for $\Delta=0.1$ $\&$ $\gamma=0.4$\\
\hline
$\hspace{0.3cm}B_{1}$ & \hspace{0.3cm}$-2$ & $-6\gamma+\Delta+2$ & $\hspace{0.5cm}7$ & \hspace{0.6cm}$1$ & saddle for $\Delta=0.1$ $\&$ $\gamma=0.7$\\[1pt]
\hline
$\hspace{0.3cm}C_{1}$ & \hspace{0.3cm}$-4$ & $-6\gamma+\Delta-8$ & $\hspace{0.3cm}-3$ & \hspace{0.3cm}$-3$ & stable for $\gamma=-0.63$, $\&$ $\Delta=0.1$\\
\hline
$\hspace{0.3cm}D_{1}$ & $-2+\frac{3z}{4\alpha}$ & $-6\gamma+\Delta+2$ & \hspace{0.5cm}$7$ & \hspace{0.6cm}$1$ & saddle for $z>\frac{4\alpha}{3}$\\[1pt]
\hline
$\hspace{0.3cm}E_{1}$ & \hspace{0.3cm}$-4$ & $-6\gamma+\Delta-8-\frac{9\beta r}{2\alpha}$ & \hspace{0.3cm}$-3$ & $-9+\frac{3\beta r}{2\alpha}$ & stable for $\frac{2\alpha(\Delta-8-6\gamma)}{9\beta}$$<r<\frac{6\alpha}{\beta}$\\[1pt]
\hline
$\hspace{0.3cm}F_{1}$ & \hspace{0.3cm}$-4$ & $-6\gamma+\Delta-8-\frac{3z}{\alpha}$ & \hspace{0.3cm}$-3$ & \hspace{0.3cm}$-3$ & stable for $z>$ $\frac{\alpha(\Delta-8-6\gamma)}{3}$\\[1pt]
\hline
$\hspace{0.3cm}G_{1}$ & $-4-\frac{3r\beta}{2\alpha}$ & $-6\gamma+\Delta+1-\frac{3\beta r}{\alpha}$ & \hspace{0.5cm}$6$ & \hspace{0.4cm}$\frac{3\beta r}{2\alpha}$ & unstable for $\frac{8\alpha}{3\beta}<$ $r<\frac{\alpha(\Delta+1-6\gamma)}{3\beta}$\\[1pt]
\hline\hline
\end{tabular}
\label{Tab:T5}
\end{table}
\begin{figure}[hbt!]
    \centering
    \begin{subfigure}[b]{0.3\textwidth}
        \includegraphics[scale=0.4]{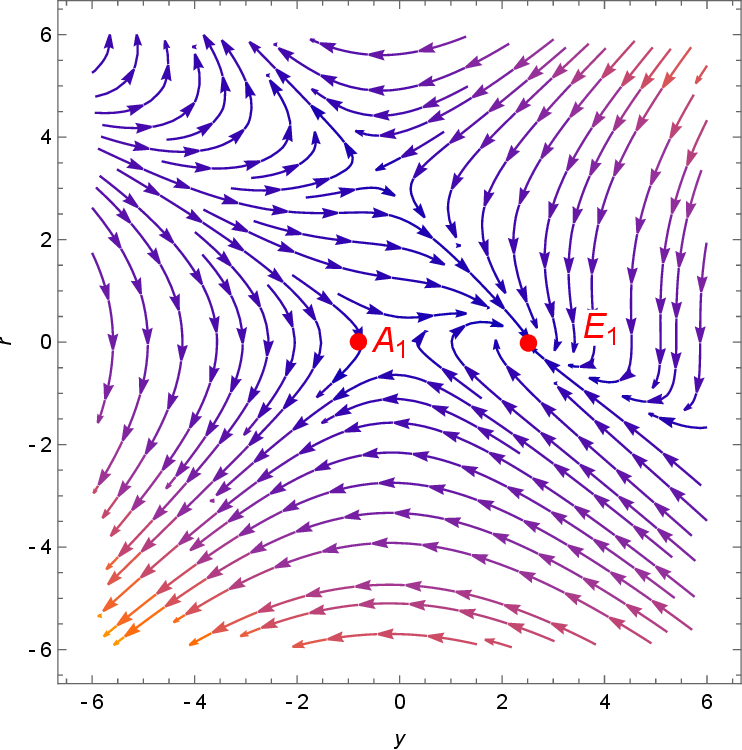}
        \caption{}
        \label{fig:subfig4}
    \end{subfigure}
    \hfill
    \begin{subfigure}[b]{0.3\textwidth}
        \includegraphics[scale=0.4]{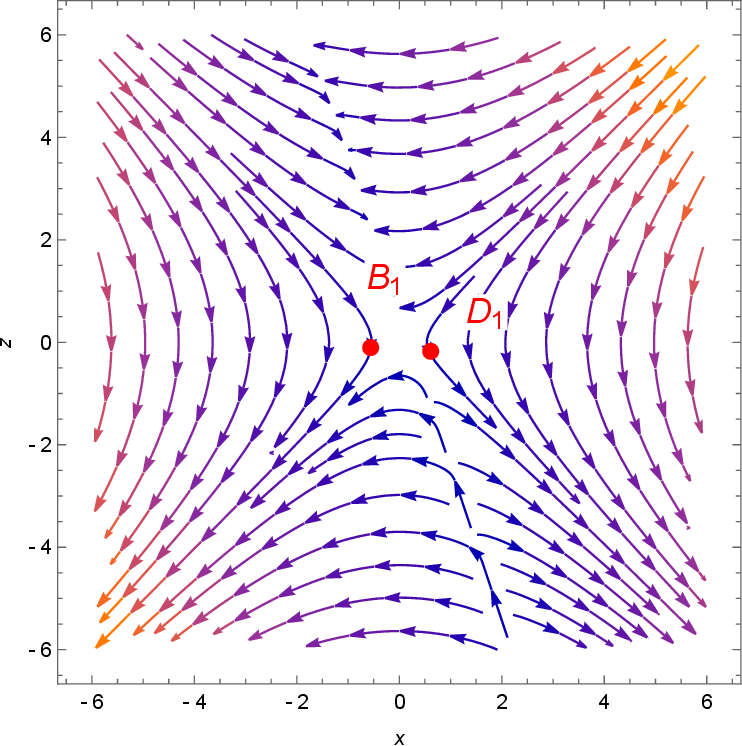}
        \caption{}
        \label{fig:subfig5}
    \end{subfigure}
    \hfill
    \begin{subfigure}[b]{0.3\textwidth}
        \includegraphics[scale=0.4]{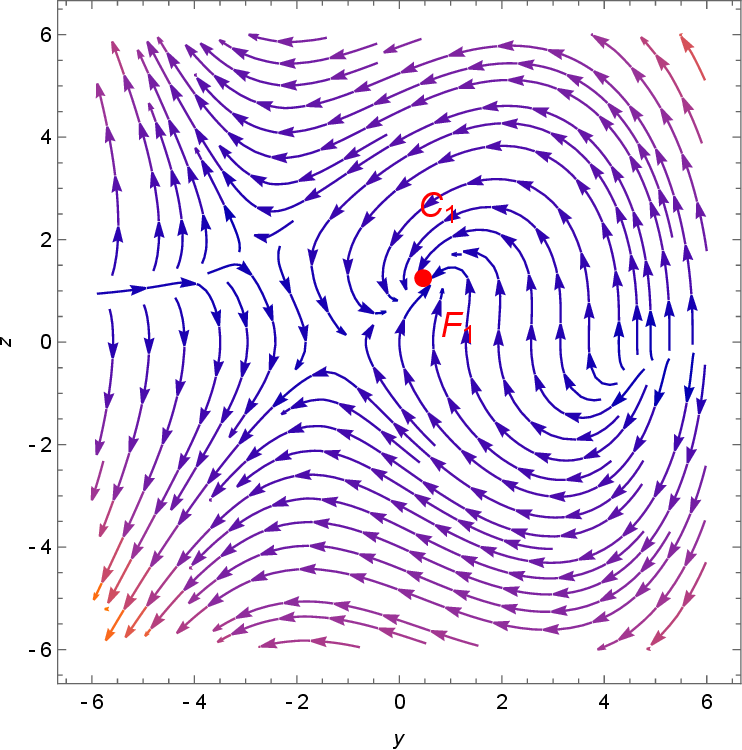}
        \caption{}
        \label{fig:subfig6}
    \end{subfigure}\\
     \vfil
     \begin{subfigure}[b]{0.3\textwidth}
        \includegraphics[scale=0.4]{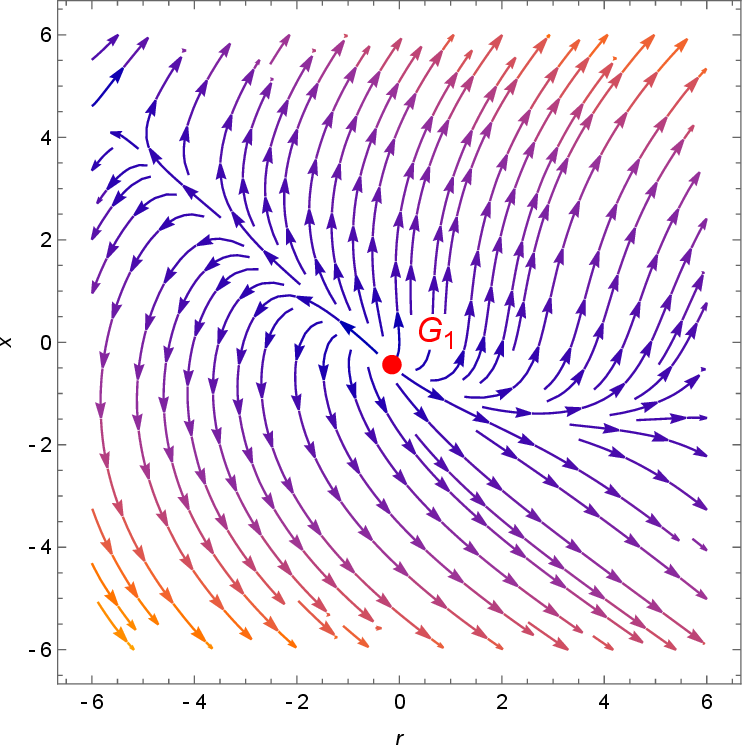}
        \caption{}
        \label{fig:subfig7}
    \end{subfigure}
    \caption{Phase space diagram for the equations (\ref{40})â€“(\ref{43}) with $\alpha=-1$, $\beta=2$ and $\Delta=0.1$.}
    \label{fig:f2}
\end{figure}
\begin{table}[h!]
\centering
\caption{Physical behavior of the scale factor at each equilibrium points}
\begin{tabular}{||p{0.9cm}|p{3.5cm}|p{3.0cm}|p{3.3cm}||}
\hline\hline
 Points & Accelerating equation & \hspace{0.4cm}Scale factor & Phase \\
\hline\hline
$\hspace{0.3cm}A_{1}$ & \hspace{0.6cm}$\dot{H}=-\frac{3}{2}H^{2}$ & $a(t)=t_{0}(\frac{3t}{2}+b_{1})^{\frac{2}{3}}$ & matter dominated\\
\hline
$\hspace{0.3cm}B_{1}$ & \hspace{0.6cm}$\dot{H}=-2H^{2}$ & $a(t)=t_{0}(2t+b_{2})^{\frac{1}{2}}$ & radiation dominated\\
\hline
$\hspace{0.3cm}C_{1}$ & \hspace{0.8cm}$\dot{H}=0$ & \hspace{0.5cm}$a(t)=t_{0}e^{b_{3}t}$ & de-Sitter\\
\hline
$\hspace{0.3cm}D_{1}$ & \hspace{0.6cm}$\dot{H}=-2H^{2}$ & $a(t)=t_{0}(2t+b_{2})^{\frac{1}{2}}$ & radiation dominated\\
\hline
$\hspace{0.3cm}E_{1}$ & \hspace{0.8cm}$\dot{H}=0$ & \hspace{0.5cm}$a(t)=t_{0}e^{b_{3}t}$ & de-Sitter\\
\hline
$\hspace{0.3cm}F_{1}$ & \hspace{0.8cm}$\dot{H}=0$ & \hspace{0.5cm}$a(t)=t_{0}e^{b_{3}t}$ & de-Sitter\\
\hline
$\hspace{0.3cm}G_{1}$ & \hspace{0.6cm}$\dot{H}=-\frac{3}{2}H^{2}$ & $a(t)=t_{0}(\frac{3t}{2}+b_{1})^{\frac{2}{3}}$ & matter dominated\\
\hline\hline
\end{tabular}
\label{Tab:T6}
\end{table}\\

$\star$ \textbf{Critical Point $A_{1}$:} The values of the critical point $A_{1}$ are $\big(0,0,0,-\frac{2\alpha}{\beta}\big)$, where $\beta\neq0$. The following eigenvalues are connected to the point $A_{1}$: $\lambda_{1}=-1$: indicates stability along one direction, $\lambda_{2}=-6\gamma+\Delta+7$: this eigenvalue will depend on the specific values of $\gamma$ and $\Delta$. For $\gamma=0.4$ and $\Delta=0.1$ showing a positive value indicating instability in one direction. $\lambda_{3}=6$: Positive, indicating instability along another direction. $\lambda_{4}=-3$: Negative, indicating stability along one direction. Since the eigenvalues have both positive and negative values, $A_{1}$ is classified as a saddle point. The Universe is attracted to this point in some directions but repelled in others.

At this point, $q=\frac{1}{2}$, which is characteristic of a matter-dominated Universe. This value of $q$ implies that the Universe is decelerating in its expansion. The value $\omega_{BD}=0.033$ is close to zero, suggesting that if BADE were present, it would behave similarly to a pressureless component. However, since $y=0$ at this point, BADE does not contribute to the dynamics. The critical point $A_{1}$ leads to $\Omega_{r}=\Omega_{BD}=0$ and $\Omega_{m}=1$, indicating a phase where the Universe is entirely matter-dominated. The scale factor evolves as $a(t)=t_{0}(\frac{3t}{2}+b_{1})^{\frac{2}{3}}$, typical of a matter-dominated Universe, reflecting the expected decelerating expansion during this epoch.

A plot \ref{fig:subfig4} of the saddle behavior around $A_{1}$ would depict trajectories converging towards this critical point in certain directions (indicating stability) and diverging in others (indicating instability). This visual representation reflects how the Universe might briefly pass through a matter-dominated phase before transitioning to another evolutionary stage.

$\star$ \textbf{Critical Point $B_{1}$:} The eigenvalues for the critical point $B_{1}$ are: $\lambda_{1}=-2$: Negative, indicating stability along one direction. $\lambda_{2}=-6\gamma+\Delta+2$: For $\gamma=0.7$ and $\Delta=0.1$, this value becomes $\lambda_{2}$$\approx$$-2.1$ indicating stability in another direction. $\lambda_{3}=7$: Positive, indicating instability in one direction. $\lambda_{4}=1$: Positive, indicating instability along another direction. Thus, $B_{1}$ is categorized as a saddle point due to the combination of positive and negative eigenvalues. The saddle behavior of $B_{1}$ is indicative of a transitional phase in the Universe's evolution. The Universe may start in a radiation-dominated era, as expected in the early stages after the Big Bang, but will eventually move away from this state due to the unstable directions indicated by the positive eigenvalues.

Here, $q=1$, which corresponds to a radiation-dominated Universe. This indicates a high deceleration rate typical of the early Universe when radiation was the dominant energy component. At the point $B_{1}$, $\omega_{BD}=-0.33$, which would usually indicate a negative pressure component typical of dark energy. The critical point $B_{1}$ leads to $\Omega_{r}=1$, $\Omega_{BD}=0=\Omega_{m}$, clearly indicating a phase where the Universe is entirely dominated by radiation. The scaling factor changes over time as $a(t)=t_{0}(2t+b_{2})^{\frac{1}{2}}$, characteristic of a radiation-dominated Universe. This is consistent with the expected expansion behavior during the radiation era, where the scale factor grows as $t^{\frac{1}{2}}$, indicating a rapid but decelerating expansion.

Plotting the saddle behaviour around $B_{1}$ as a \ref{fig:subfig5} would display trajectories diverging in some directions, indicating instability, and converging in others, indicating stability, towards this critical point.

$\star$ \textbf{Critical Point $C_{1}$:} The eigenvalues for the critical point $C_{1}$ are: $\lambda_{1}=-4$: Negative, indicating stability along one direction. $\lambda_{2}=-6\gamma+\Delta-8$: For $\gamma=-0.63$ and $\Delta=0.1$, this gives $\lambda_{2}\approx-4.6$ indicating stability in another direction. $\lambda_{3}=-3=\lambda_{4}$: Negative, indicating stability in the another direction. With all eigenvalues being negative, $C_{1}$ is classified as stable point. This means that the Universe will remain in this accelerated, dark energy-dominated phase indefinitely, leading to a stable de Sitter expansion.

For the point $C_{1}$, $q=-1$, which corresponds to an accelerated expansion of the Universe, characteristic of a de Sitter phase and the EoS parameter $\omega_{BD}=\frac{-\Delta-1}{3}$. For $\Delta=0.1$, this gives $\omega_{BD}\approx-0.367$. This value is typical of a dark energy equation of state and indicates that BADE is acting as the primary driver of the accelerated expansion at this critical point. The density parameters are $\Omega_{m}=\Omega_{r}=0$ and $\Omega_{BD}=1$, indicating that the Universe is entirely dominated by BADE at this point. The scaling factor changes over time as $a(t)=t_{0}e^{b_{3}t}$, characteristic of an exponential expansion typical of a de Sitter universe. This indicates that the Universe is in a phase of constant, accelerated expansion, which is expected in a scenario dominated by dark energy with a negative pressure.

A plot \ref{fig:subfig6} illustrating the stability around $C_{1}$ would display trajectories converging toward this critical point from every direction, signifying that the Universe will inevitably transition into this dark energy-dominated phase.

$\star$ \textbf{Critical Point $D_{1}$:} The eigenvalues associated with this critical point are: $\lambda_{1}=-2+\frac{3z}{4\alpha}$: For $z>\frac{4\alpha}{3}$, this eigenvalue becomes positive, indicating instability in one direction. $\lambda_{2}=-6\gamma+\Delta+2$: This eigenvalue's behavior depends on the specific values of $\gamma$ and $\Delta$. For $\gamma=0.7$ and $\Delta=0.1$, this value becomes $\lambda_{2}$$\approx$$-2.1$ indicating stability in another direction. $\lambda_{3}=7$: Positive, indicating instability in another direction. $\lambda_{4}=1$: Positive, also indicating instability in the final direction. With a mix of positive and negative eigenvalues, $D_{1}$ is classified as a saddle point. The saddle nature of $D_{1}$ indicates that while the Universe may spend some time in this radiation-dominated phase, it is not a final state. The positive eigenvalues suggest that the Universe will eventually transition out of this phase, moving toward a matter-dominated phase or, eventually, a dark energy-dominated phase as seen in later cosmological epochs.

The deceleration parameter $q=1$ is indicative of a Universe in a decelerating expansion phase, characteristic of radiation-dominated eras. This is consistent with the standard cosmological model during the early Universe, where radiation led to a decelerating expansion. The number $\omega_{BD}=-0.33$, which ordinarily denotes a dark energy-typical negative pressure component. The density parameters are: $\Omega_{r}=1$, $\Omega_{BD}=0=\Omega_{m}$, representing the radiation-dominated phase of the Universe. The scale factor $a(t)=t_{0}(2t+b_{2})^{\frac{1}{2}}$ describes a Universe whose expansion is governed by radiation. This form of the scale factor is consistent with the radiation-dominated era, where $a(t)\propto t^{\frac{1}{2}}$. This implies that the Universe expands at a rate that decreases over time, which is typical in a radiation-dominated epoch.

The plot \ref{fig:subfig5} shows trajectories that approach $D_{1}$, representing different possible states of the Universe that are consistent with radiation dominance. However, due to the saddle nature of this point, these trajectories eventually diverge, highlighting the fact that the Universe is compelled to evolve beyond this phase.

$\star$ \textbf{Critical Point $E_{1}$:} The critical point $E_{1}$ exists for $r\neq0$ and within specific parameter ranges, specifically $\beta\neq 0$, $0<\Delta<1$. The eigenvalues at $E_{1}$ are: $-4,-6\gamma+\Delta-8-\frac{9\beta r}{2\alpha},-3,-9+\frac{3\beta r}{2\alpha}$. These eigenvalues indicate that $E_{1}$ can be a stable point, provided the condition $\frac{2\alpha(\Delta-8-6\gamma)}{9\beta}$$<r<\frac{6\alpha}{\beta}$ is satisfied.

The stability of $E_{1}$ implies that the Universe can settle into this state, leading to sustained accelerated expansion, as indicated by the deceleration parameter $q=-1$. The equation of state parameter $\omega_{BD}=\frac{-\Delta-1}{3}$ is negative, reinforcing the interpretation of a dark energy-dominated phase where the Universe experiences accelerated expansion. The density parameters $\Omega_{m}=\Omega_{r}=0$ and $\Omega_{BD}=1$ suggest that the point $E_{1}$ corresponds to a scenario dominated entirely by the BADE. The complete dominance of BADE at this critical point suggests that it plays a crucial role in driving the late-time accelerated expansion of the Universe. The negative equation of state parameter, coupled with the stability of this point, implies that BADE acts as a form of dark energy that could explain the observed accelerated expansion of the Universe. The parameters $\Delta$ and $\gamma$ control the behavior of BADE, with their values influencing the stability and the range of $r$. This sensitivity to the parameters highlights the model's ability to describe various cosmological scenarios, depending on the specific values chosen.

The scale factor at $E_{1}$ is given by $a(t)=t_{0}e^{b_{3}t}$, which corresponds to an exponentially expanding Universe with the expansion rate $b_{3}$ determined by the underlying parameters $\alpha$, $\beta$, $\Delta$ and $\gamma$. This exponential expansion is a hallmark of dark energy dominance of the Universe.

The stability plot \ref{fig:subfig4} for $E_{1}$ would likely show trajectories converging to this point, indicating that once the Universe reaches this state, it remains there. This convergence is consistent with the stable nature of $E_{1}$, suggesting that the Universe's expansion rate becomes constant, leading to a phase where the dark energy density dominates and drives the expansion.

$\star$ \textbf{Critical Point $F_{1}$:} The critical point $F_{1}$ exists under the conditions $z\neq 0$, $\beta\neq 0$, $0<\Delta<1$. The expression $\frac{1}{\Delta+4-3\gamma}\bigg[-\frac{12\alpha}{\beta}+\frac{3z}{\beta}\bigg]$ suggests a complex interplay between the parameters $\alpha$, $\beta$, $\Delta$ and $\gamma$, which determines the exact position of the critical point. The eigenvalues $-4,-6\gamma+\Delta-8-\frac{3z}{\alpha},-3,-3$ suggest that the point $F_{1}$ is a stable critical point provided the condition $z>$ $\frac{\alpha(\Delta-8-6\gamma)}{3}$ is met. This condition emphasizes that the stability of $F_{1}$ is controlled by the magnitude of $z$, linking the dynamic aspects of dark energy directly to the stability of the cosmological model. The stability of $F_{1}$ means that, once the Universe reaches this state, it remains there indefinitely, implying that $F_{1}$ could represent the final attractor in the cosmic evolution. The presence of the parameter $z$ in the critical point's definition introduces a degree of freedom that can alter the dark energy dynamics. The condition $z>$ $\frac{\alpha(\Delta-8-6\gamma)}{3}$ indicates that the dark energy model described by $F_{1}$ is not static but allows for a dynamic evolution depending on $z$.

The deceleration parameter $q=-1$ and the EoS parameter $\omega_{BD}=\frac{-\Delta-1}{3}$ are negative, which reinforces the interpretation that BADE behaves like dark energy, causing the accelerated expansion of the Universe. The density parameters $\Omega_{m}=\Omega_{r}=0$ and $\Omega_{BD}=1$ confirm that $F_{1}$ is a state of complete dark energy dominance. The scale factor $a(t)=t_{0}e^{b_{3}t}$ at $F_{1}$ represents exponential expansion, characteristic of a de Sitter-like Universe.

The stability plot \ref{fig:subfig6} for $F_{1}$ would show trajectories converging to this point. This behavior suggests that $F_{1}$ acts as a cosmic attractor, with the Universe settling into this state where dark energy dominates the dynamics.

$\star$ \textbf{Critical Point $G_{1}$:} The critical point $G_{1}$ exists for $r\neq 0$, $\beta\neq 0$. The expression $-\frac{3\alpha}{\beta}-\frac{3r}{2}$ for the first critical coordinate indicates that the point is significantly influenced by the parameters $\alpha$, $\beta$ and $r$. This dependence shows that the dynamics at $G_{1}$ are tied to the interaction between the model's parameters and the specific choice of $r$, which could correspond to a certain interaction term or a correction in the dark energy sector. The eigenvalues $-4-\frac{3r\beta}{2\alpha}, -6\gamma+\Delta+1-\frac{3\beta r}{\alpha},6,\frac{3\beta r}{2\alpha}$ indicate that $G_{1}$ is an unstable critical point within the specified range of $r$. Specifically, the positive eigenvalue $6$ and the condition $\frac{8\alpha}{3\beta}<$ $r<\frac{\alpha(\Delta+1-6\gamma)}{3\beta}$ ensure instability, leading to divergence away from $G_{1}$. The instability at $G_{1}$ implies that this point cannot represent a final or stable phase in the cosmic evolution. Instead, it likely corresponds to a transient phase where the Universe temporarily passes through a matter-dominated era before moving towards another attractor or stable state.

The deceleration parameter $q=\frac{1}{2}$ at $G_{1}$ is characteristic of a matter-dominated era, where the Universe is still decelerating due to gravitational attraction. This value of $q$ aligns with the standard cosmological model during the matter-dominated epoch, where the scale factor evolves as $a(t)=t_{0}(\frac{3t}{2}+b_{1})^{\frac{2}{3}}$. The presence of the constant $b_{1}$ may indicate an initial condition or a correction term related to the onset of matter domination. The equation of state parameter $\omega_{BD}=0.033$, close to zero, indicates that the BADE component behaves similarly to a matter-like substance at this point. The small positive value suggests that BADE provides a slight pressure, but its effect is negligible compared to the dominant matter component, represented by $\Omega_{m}=1$. The density parameters $\Omega_{r}=\Omega_{BD}=0$ and $\Omega_{m}=1$ further emphasize the matter-dominated nature of $G_{1}$. At this point, radiation and BADE contribute negligibly, leaving matter as the sole component driving the expansion of the Universe.

The instability plot \ref{fig:subfig7} for $G_{1}$ would show trajectories diverging from this point, reflecting its transient nature. As the Universe evolves, it would move away from $G_{1}$, indicating that this point does not represent a long-term solution in the phase space of the dynamical system.

Observationally, the existence of $G_{1}$ may relate to epochs where matter's influence was paramount, and dark energy was subdominant. As the Universe evolves, moving away from $G_{1}$, it could naturally transition to a state that aligns with current observations of an accelerating Universe.
\section{Conclusion}\label{sec4}
\hspace{0.5cm} In this analysis, we explored the dynamics of the BADE model within the framework of $f(Q,L_{m})$ gravity by considering both non-interacting and interacting scenarios. The critical points derived from the dynamical system analysis reveal distinct cosmological phases that align with the theoretical expectations for the evolution of the Universe, from matter dominance to a dark energy-dominated accelerated expansion.

$\bullet$ \textbf{Non-interacting case:} The non-interacting scenario highlighted several critical points where the Universe could experience different phases, including matter domination, radiation domination, and a BADE-dominated accelerated expansion. Among these, two stable critical points were identified, corresponding to a BADE-dominated phase with accelerated expansion. These stable points suggest that the BADE model within $f(Q,L_{m})$ gravity can naturally explain the transition from a decelerated matter-dominated phase to an accelerated dark energy-dominated phase. The existence of stable attractors with negative deceleration parameters $(q<0)$ and negative equations of state $(\omega_{BD}<-\frac{1}{3})$ supports the potential of the BADE model to drive the late-time acceleration of the Universe. These results align with current observational data, which indicate a transition to accelerated expansion in the current epoch.

$\bullet$ \textbf{Interacting case:} In the interacting scenario, where energy exchange occurs between BADE and matter, the dynamical system analysis revealed more complex behaviors, including the possibility of unstable and saddle points that represent transient phases in cosmic evolution. Crucially, in the interaction scenario, three stable critical points were identified. These stable points are especially important because they suggest that the Universe will eventually settle into a BADE-dominated period of accelerated expansion, following potentially brief times dominated by matter. At the stable points, the deceleration parameter $q=-1$, indicating a phase of accelerated expansion. Additionally, the equation of state parameter for Barrow dark energy, $\omega_{BD}=\frac{-\Delta-1}{3}$, is negative, showing that BADE behaves like a dark energy component capable of driving the accelerated expansion. The analysis shows that interaction between BADE and matter can lead to different cosmological behaviors, but the Universe ultimately converges to a stable accelerated phase characterized by a consistent deceleration parameter and a negative equation of state. It can be inferred from this that the interaction facilitates a more complex evolution of the Universe's dynamics rather than preventing the shift towards accelerated expansion.

The results highlight the significance of the interaction term and its impact on the stability and evolution of the Universe's critical points. Finally, the dynamical system approach used in this analysis provides valuable insights into the stability and behavior of cosmological models, confirming the idea that $f(Q,L_{m})$ gravity with BADE is a promising candidate for describing the late-time evolution of the Universe.                                                                                               
 
 \end{document}